\documentclass[12pt,a4paper]{article}

\usepackage{epsfig} 
\usepackage{graphicx}

\begin{document}
\textwidth=135mm
 \textheight=200mm
\begin{center}
{\bfseries Prospects of the search for neutrino bursts from Supernovae with Baksan Large Volume Scintillation Detector 
\footnote{{\small Talk at the International Workshop on Prospects of Particle Physics: 
"Neutrino Physics and Astrophysics" , Valday, Russia, January 1 - February 8, 2015.}}}
\vskip 5mm
V. B. Petkov$^{\dag,\ddag}$
\vskip 5mm
{\small {\it $^\dag$ Institute for Nuclear Research of RAS, Moscow, Russia}} \\
{\small {\it $^\ddag$ Institute of Astronomy of  RAS, Moscow, Russia}}
\\
\end{center}
\vskip 5mm
\centerline{\bf Abstract}
Observing a high-statistics neutrino signal from the supernova explosions in the Galaxy is a major goal of low-energy neutrino astronomy. The prospects for detecting all flavors of neutrinos and antineutrinos from the core-collapse supernova (ccSN) in operating and forthcoming large liquid scintillation detectors (LLSD) are widely discussed now. 
One of proposed LLSD is Baksan Large Volume Scintillation Detector (BLVSD). This detector will be installed at the Baksan Neutrino Observatory (BNO) of the Institute for Nuclear Research, Russian Academy of Sciences, at a depth of 4800 m.w.e. Low-energy neutrino astronomy is one of the main lines of research of the BLVSD.

\vskip 10mm

\section{\label{sec:intro}Introduction}
The experiment to search for neutrino bursts from supernovae is carried out at the Baksan Underground 
Scintillation Telescope (BUST) for many years. It is the longest duration experiment (June 1980 - December 2014, $T_{live}$ = 29.76 years) with best upper limit on the rate of core-collapse and failed supernova explosions in the Galaxy: $f < 0.077$ $y^{-1}$ at 90\% confidence level \cite{Novoseltseva15}. 
The BUST, being the oldest of neutrino telescopes in the world, has the total scintillator mass of 330 tons. It allows one to detect the neutrinos from the ccSN in the Galaxy but insufficient to study of details of neutrino burst.

A large volume detector filled with liquid scintillator at the Baksan neutrino observatory is discussed for long time \cite{Domogat1, Domogat2, Domogat3, Domogat4, Barabanov1}. The main research activities of the BLVSD are neutrino geophysics and neutrino astrophysics. 
At present R$\&$D work aimed at the creation of a new-generation detector using an extra-pure scintillator of 5 - 20 kiloton mass is performed \cite{Lubsand1, Bezrukov1, Barabanov2, Barabanov3}.

Because the supernova explosion in the Galaxy is a rare event, the comprehensive study of the next one has absolute priority for the astronomy, including low-energy neutrino astronomy.
This circumstance imposes strict requirements on the new-generation LLSDs.

It is essential be ready to detect all flavors of neutrinos in order to understand the physics and astrophysics of core-collapse supernovae \cite{Lujan, Laha14}.
For this purpose the new detector must have the capability to distinguish the various detection channels.
Large statistics must be collected to study spectra and time profiles of all  neutrino flavors. Therefore the new detector should have large enough target mass.

The optical/near-IR observations will remain a crucial component
of studies of Galactic ccSNe. Since neutrinos precede the optical photons by several hours, early ccSN detection and location by neutrinos is of primary importance \cite{Adams}. It could allow observation of the evolution of the first optical stages to be made. 
To this effect the new detector must give alert, time, and direction as rapidly as possible.
Below some methods of determination of ccSN position will be considered.

\section{Determination of direction to ccSN}
In water Cerenkov detectors (WCD) the neutrino-electron scattering:
\begin{equation}\label{ES1}
 \nu_{e} + e^{-} \to \nu_{e} + e^{-},
\end{equation} 
where $i = e, \mu, \tau$, can be used for location of a Galactic supernova. These events are forward-peaked, so a narrow cone contains the majority of them. Neutrino-electron forward scattering leads to a good determination of the supernova direction, in spite of the large and nearly isotropic background from other reactions. Even with the most pessimistic background assumptions, SuperKamiokande can restrict the supernova direction to be within circles of a radius of $5^{\circ}$ \cite{Beacom99}.

The background from other reactions can be reduced by means of recognizing  electron neutrinos from the short-lived neutronization burst of $\nu_e$ events
from a supernova. Prior to the core explosion, many $\nu_e$'s are emitted via the reaction $e^{-} + p \to \nu_{e} + n$ as the shock wave propagates into the exploding stars outer core. The shock wave dissociates nuclei into free nucleons, for which the cross section of electron
capture is larger than that for nuclei; the resulting burst of $\nu_e$'s thus
forms the so-called neutronization burst. The duration of the neutronization burst is on the timescale of the shockwave propagation, which is less than 10 ms \cite{SK07}.

Unlike WCD large liquid scintillation detectors do not provide direct angular information, except for charged leptons in the GeV range with track lengths greater than the typical resolution of LLSDs of tens of centimeters \cite{Wurm}.
Nevertheless, indirectly one can retrieve directional information using inverse beta decay (IBD) reaction:
 \begin{equation}\label{IBD}
 \bar{\nu}_{e} + p \to n + e^{+}.
\end{equation} 
The high cross section of this reaction leads to a high number of interactions and the statistical analysis of the angular distributions of the reaction products can be correlated to the neutrino direction.
The delayed neutron capture by a proton is characterized by a monochromatic gamma 2.2 MeV emission. The coincidence in a typical time window of about 250 $\mu$s between the latter and the prompt signal from the $e^+$ gives a clear signature of an IBD event.
In the IBD reaction the direction of the outgoing positron is nearly isotropic because of the small recoil energy of a nucleon. 
But the angular distribution of the outgoing neutrons is strongly forward-peaked, leading to a measurable separation in positron and neutron detection points \cite{Vogel99}. The offset between the $e^+$ and the neutron-capture locations can be reconstructed, although with large uncertainties. 
A directional sensitivity of this technique have been demonstrated in the CHOOZ reactor neutrino experiment, in which reactor location was reconstructed
within $18^{\circ}$ half-cone aperture (68\% C.L.) on the basis of 2500 reconstructed events in a Gd-loaded scintillator. Using the ccSN $\bar{\nu}_{e}$ energy distribution of 5000 neutrino interactions was generated in an experiment with the same geometry, the same position resolution and the same target (Gd-loaded liquid scintillator) as the CHOOZ experiment. The resulting uncertainty in the direction measurement is $8.8^{\circ}$ for all events and $8.4^{\circ}$ for events with positron-neutron distance larger than 20 cm \cite{Apollonio}. 

It should be noted that the use of this technique needs very large scintillation detectors, as in currently considered projects JUNO (20 kt) or LENA (50 kt). The point is that a single detector of 1 kton mass can measure the supernova position with appropriate accuracy, if the distance to the supernova doesn't exceed a few kpc \cite{Fischer}. 
On the other hand, the condition for a successful IBD tag is that no more than one interaction occurs during the time between the IBD interaction and the neutron capture inside a specific volume. This condition gives a limitation on the minimal distance D to the supernova depending on mass of the detector. While for the detector with a mass of 1 kton  $D\ge 0.16$ kpc, for similar detector with a 50 kton mass the minimal distance becomes $D\ge 1.13$ kpc, that includes several known potential Core-Collapse Supernovae \cite{Lujan}.

\section{New techniques}

The recent development of new experimental techniques has opened up the possibility for a new kind of large-scale detectors capable both to detect all flavors of neutrinos and to reconstruct the supernova location.  

First of all, one should mention the Advanced Scintillation Detector Concept (ASDC), which combines the use of Water-based Liquid Scintillator (WbLS) with high efficiency and ultra-fast timing photosensors. WbLS uniquely offers the high light yield and low threshold of scintillator with the directionality of a Cherenkov detector. The use of high-precision timing measurement devices would allow one to separate the prompt Cherenkov light from the delayed scintillation.
So, the ASDC combines the benefits of both, water Cherenkov detector and pure liquid scintillation detector,
in a single detector (see \cite{Alonso14} and references therein).

Another approach exploits a glowing track of charged particle.
A charged particle traversing a liquid scintillator
induces scintillation along its track. At each
point of the track the produced light is emitted
isotropically. The use of suitable multi-pixel photodetectors (such as CCD or SiPM matrices) with appropriate optical collector gives, in principle, a possibility to do a snapshot of this glowing track.
This technique has the obvious advantages. Firstly, the snapshot of glowing track of the particle gives a possibility in principle to determine the direction of the  particle.
Secondly, there is a possibility to measure the energy release along the track of particle. 

This technique is under development now in INR RAS. The SiPM matrices are used as multi-pixel photodetectors in the prototype of such scintillation detector (Fig. 1). 

\begin{figure}[tph!]
\centerline{\psfig{file=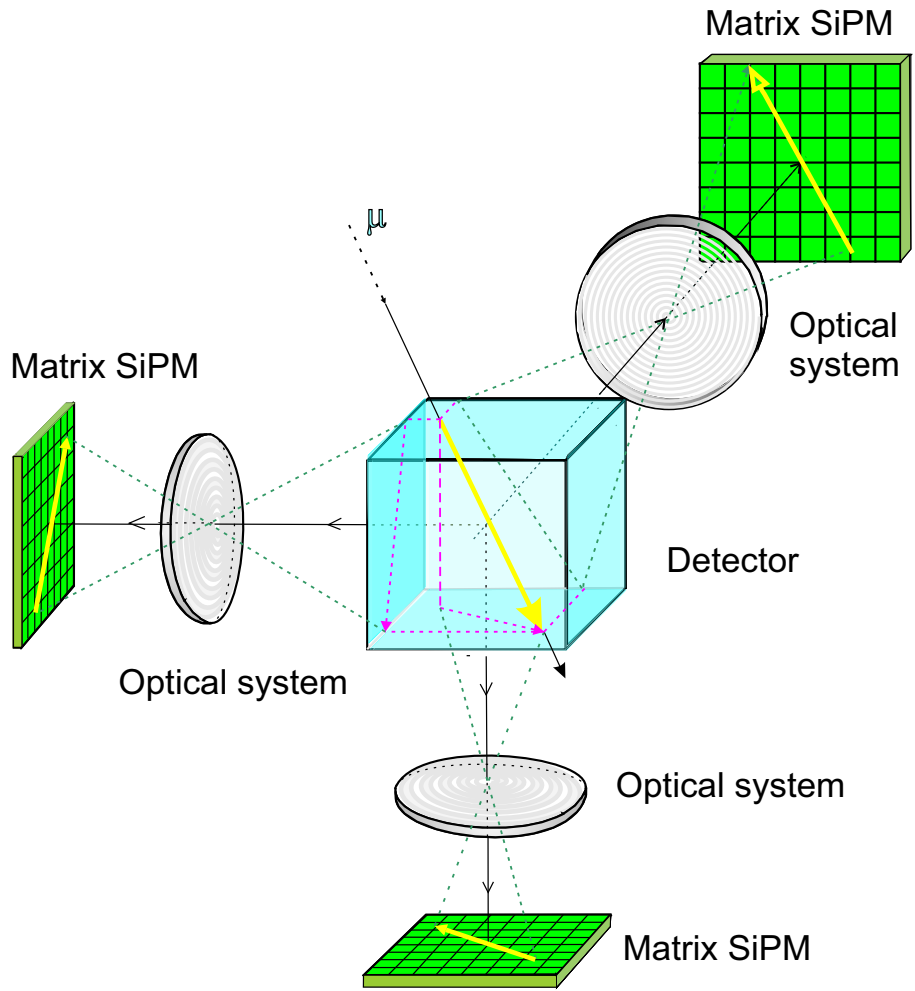,width=13cm}}
\caption{A sketch of the prototype of scintillation detector with the SiPM matrices as multi-pixel photodetectors.}
\end{figure}

Indeed, this technique is not easy to realize in LLSD. It may be shown by means of simple estimation. 
Let $\Delta x$ is minimal lenght of track of the particle, whose light comes only on one pixel of the photodetector (single SiPM in case of the SiPM matrix) from the distance $R$.
The mean number of photoelectrons $\bar{n}_{p.e.}$ in the pixel can be estimated as:
\begin{equation}
\bar{n}_{p.e.} = \bar{N}_{ph}(x) \cdot \delta (R) \cdot \Delta x \cdot k_{opt} \cdot Q_{pd} \cdot k_{scint}(R)
\end{equation} 
where $\bar{N}_{ph}(\Delta x)$ is the mean number of photons emitted on the lenght $x$ of the track; 
$$\delta (R) = \frac{A}{4\pi R^2}$$ is the part of photons arriving to the optical collector with aperture $A$ from the distance $R$;
$k_{opt}$ is the transmissivity of the optical collector; $Q_{pd}$ is the quantum efficiency of the photodetector and $k_{scint}(R) = \exp(-R/L)$ is the absorption factor in the scintillator with attenuation length $L$.

Let us consider a liquid scintillator with linear-alkyl-benzene (LAB) as solvent, with the attenuation length $L=20$ m and the light yield of 9000 photons per MeV \cite{Wurm1, Wurm2, Bezrukov1}.
Assume, for simplicity, that the energy losses of the particle is 2 MeV/(g$\cdot$ cm$^{-2}$); the transmissivity of the optical collector is $k_{opt}=0.8$ and the quantum efficiency of the photodetector is $Q_{pd}=0.4$.
In this case the mean number of photoelectrons in the pixel can be presented as:
\begin{equation}\label{MeanPh2}
\bar{n}_{p.e.} = 531.1 \cdot \Delta x \cdot A \cdot \frac{\exp(-R_{c}/L)}{R_{c}^2}
\end{equation}
(the length is measured in cm), with the distance to the center of the detector $R_c$:
$$R_c = \frac{1}{2} \cdot \sqrt[3]{\frac{M_{sc}}{\rho_{sc}}},$$
where $M_{sc}$ is the target mass of the detector and $\rho_{sc}$ is the density of the scintillator. 

Let the photodetector will be noiseless, then the number of photoelectrons on the pixel must be $\ge 1$. For LLSD with a target mass of 5 kilotons this condition is satisfied with relatively small $\Delta x$ only when the aperture of the optical collector is large enough (Fig.2). When using an optical system with the same aperture, for detecting the tracks with low path length a detector with a smaller mass is required  (Fig.3). 

Another challenge with this method is that, as opposed to PMTs, each  photodetector has a large number of channels. 
Indeed, to detect the tracks with length $\Delta x$ from the distance $R_c$ the number of channels must be much more than $A/{\Delta x}^2$ (the exact number of channels depends on specific features of optical system).

\section{Conclusion}
Obviously, that for the precise study of detected events the target mass of the new detector should not be too large.
On the other hand, large target mass of the detector is needed for obtaining large statistics of the neutrino events. 
This apparent contradiction can be resolved by creating a network of identical LLSDs, with the target mass of LLSD in the range of 2 - 5 kilotons.
This approach seems also more beneficial to study the internal structure of the Earth by geoneutrino, for this purpose the detectors of the network should be placed in sites with low background of the reactors antineutrino (see, for example, \cite{Baldoncini}). It should be noted that the Baksan Neutrino Observatory is one of optimal sites for the location of the detector of the network.

\vspace{5mm}

{\bf Acknowledgements.}\\
This work was supported by the the Russian Foundation for Basic Research  (grant 14-22-03075). 

\begin{figure}[tph!]
\centerline{\psfig{file=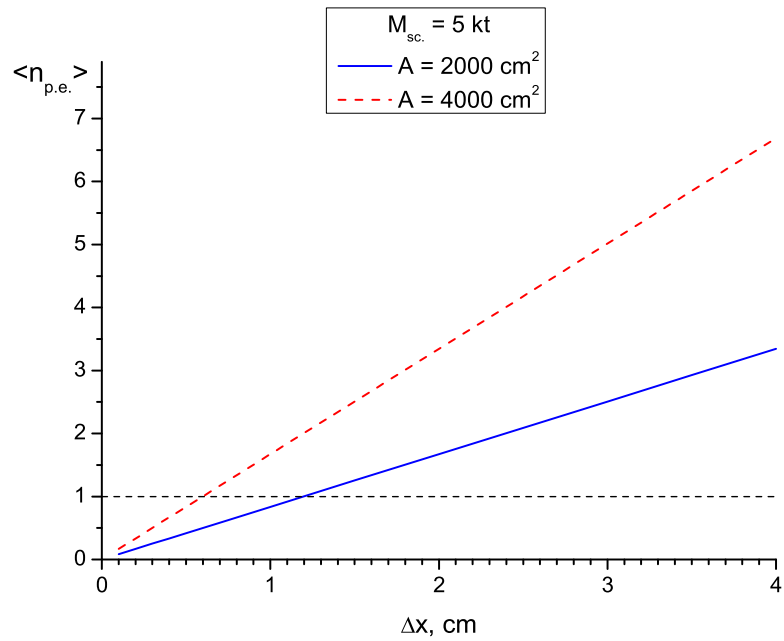,width=13cm}}
\caption{The number of photoelectrons vs. minimal length of the track for the detector with the mass of 5 kilotons.}
\end{figure}

\begin{figure}[tph!]
\centerline{\psfig{file=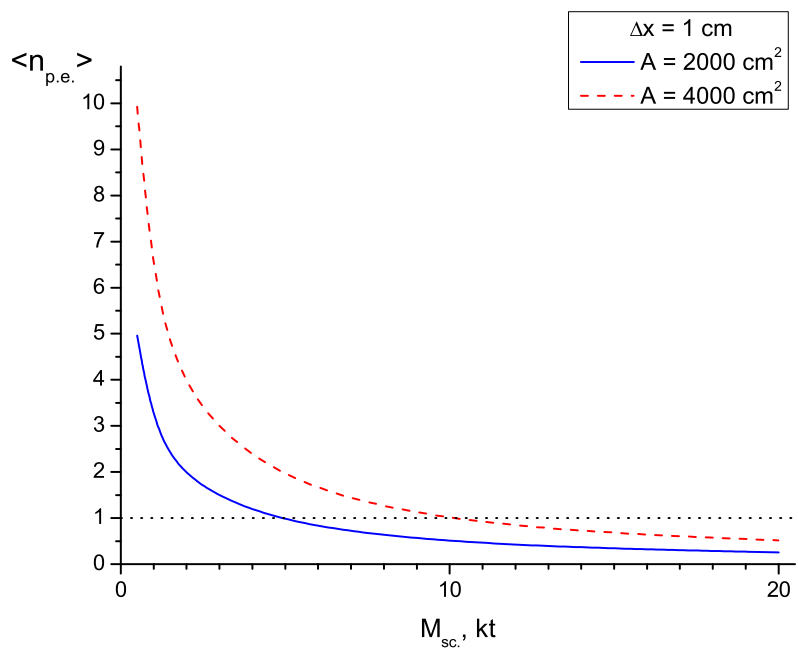,width=13cm}}
\caption{The number of photoelectrons vs. mass of the LLSD for the minimal length of the track of 1 cm.}
\end{figure}

\end{document}